\documentclass[a4paper]{article}
\usepackage[dvipsnames]{xcolor}
\definecolor{mygray}{gray}{0.6}
\usepackage{INTERSPEECH2020}
\usepackage{cite}

\title{End-to-End Multi-Look Keyword Spotting}
\name{Meng Yu$^{1}$, Xuan Ji$^{2*}$\thanks{$^*$X. Ji contributed to this work when she was with Tencent}, Bo Wu$^2$, Dan Su$^2$, Dong Yu$^1$}
\address{
  $^1$Tencent AI Lab, Bellevue, WA, USA\\
  $^2$Tencent AI Lab, Shenzhen, China}
\email{\{raymondmyu, lambowu, dansu, dyu\}@tencent.com, mrvsyzb@163.com}

\begin{document}

\maketitle
\begin{abstract}
The performance of keyword spotting (KWS), measured in false alarms and false rejects, degrades significantly under the far field and noisy conditions. In this paper, we propose a multi-look neural network modeling for speech enhancement which simultaneously steers to listen to multiple sampled look directions.
The multi-look enhancement is then jointly trained with KWS to form an end-to-end KWS model which integrates the enhanced signals from multiple look directions and leverages an attention mechanism to dynamically tune the model's attention to the reliable sources.
We demonstrate, on our large noisy and far-field evaluation sets, that the proposed approach significantly improves the KWS performance against the baseline KWS system and a recent beamformer based multi-beam KWS system.
\end{abstract}

\section{Introduction}
With the proliferation of smart homes and mobile and automotive devices, speech-based human-machine interaction becomes prevailing. To achieve hands-free speech recognition experience, the system continuously listens for specific wake-up words, a process often called keyword spotting (KWS)\cite{Rohlicek89}, to initiate speech recognition. For the privacy concern, the wake-up KWS typically happens completely on the device with low footprint and power consumption requirement.

The KWS systems usually perform well under clean-speech conditions. However, their performance degrades significantly under noisy conditions, particularly in multi-talker environments. A variety of front-end enhancement methods have been proposed in recent years, which filter out the signals of interference from the noisy stream before passing it to the KWS system. Beyond the conventional speech denoising approaches \cite{Ephraim85,Wang06} and the recent deep learning based techniques for speech enhancement \cite{Wang14, Xu15, Weninger15, Hershey16, Chen17c,Kolbk17}, a neural network based text-dependent speech enhancement technique for recovering the original clean speech signal of a specific content has been recently proposed and applied to KWS as a front-end processing component \cite{MYu18}.
However, the far field speech processing suffers from the reverberation and multiple sources of interference which blurs speech spectral cues and degrades the single-channel speech enhancement. Since the microphone array is more widely deployed than before, multi-channel techniques become more and more important. An array of microphones provides multiple recordings, which contain information indicative of the spatial origin of a sound source. When sound sources are spatially separated, with microphone array inputs one may localize sound sources and then extract the source from the target direction.


The well established spatial features, such as inter-channel phase difference (IPD), have been proven efficient when combined with monaural spectral features at the input level for time-frequency (T-F) masking based speech separation methods \cite{chen2018multi, wang2018multi, lianwu2019multi}. Furthermore, in order to enhance the source signal from a desired direction, elaborately designed directional features associated with a certain direction that indicate the directional source's dominance in each T-F bin have been presented in \cite{wang2019combining, gu2019neural,bahmaninezhad2019comprehensive}.
Nevertheless, the knowledge of the true target speaker direction is not available in real applications. It is hence very difficult to accurately estimate the target speaker's direction of arrival (DOA) in multi-talker environments. In other words, the systems don't even know which one is the target speaker among multiple acoustic sources. A multi-channel processing approach handled by fixed beamformers with multiple fixed beams has been presented in \cite{Ji20}. Instead of detecting keywords by evaluating 4 beamformed channels in consequence and indicating a successful detection if any of the 4 trials triggers the threshold, the authors developed to train a KWS model with all beamformed signals at multiple look directions as the input.
The system optimizes the multi-beam feature mapping and the keywords detection model to improve the keyword recognition accuracy.

The beamforming based multi-look approach \cite{Ji20} motivates us to work towards a neural network modeling of multi-look speech enhancement, which thus enables the joint training with KWS model to form a completely end-to-end multi-look KWS modeling. We solve the major difficulty on assigning supervised training targets to the multi-look enhancement modeling. The presented multi-look enhancement incorporates spectral features, IPDs and directional features associated with multiple sampled look directions for source enhancement in multiple directions simultaneously. It shows significant advantage compared to the conventional beamformers for the purpose of KWS with no prior information of the target speaker's location.
The rest of the paper is organized as follows. In Section \ref{sec:mlkws}, we first recap the direction-aware enhancement modeling, and then present the multi-look speech enhancement model followed by the end-to-end multi-look KWS model. We describe our experimental setups and evaluate the effectiveness of the proposed system in Section \ref{exp}. We conclude this work in Section \ref{con}.

\begin{figure*}[!t]
        \centering
        \includegraphics[width=\linewidth]{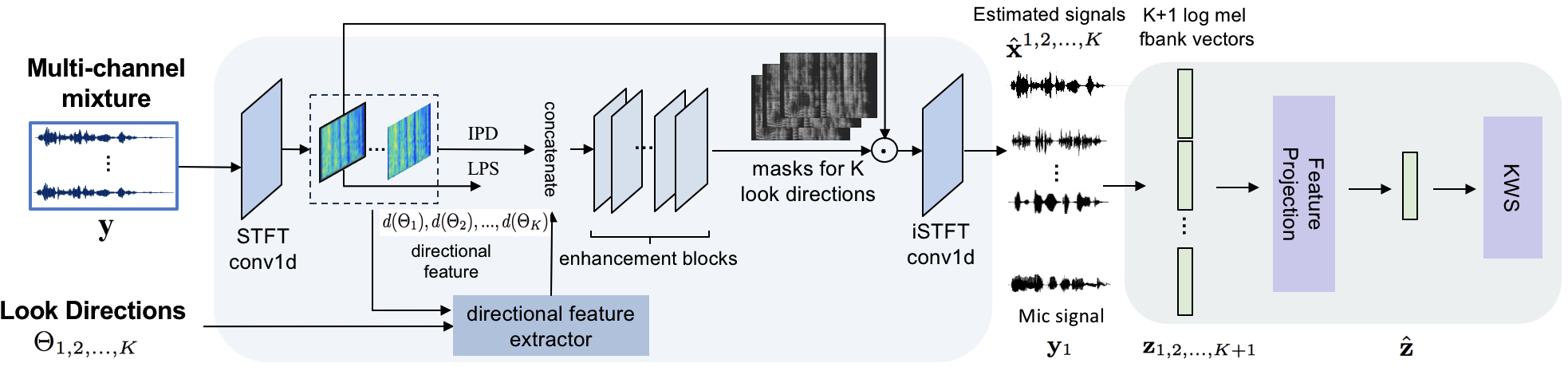}
        \caption{Block diagram of the proposed multi-look enhancement and end-to-end multi-look KWS model}
        \label{fig:mlkws}
        \vspace{-0.5cm}
\end{figure*}

\section{Multi-Look KWS}\label{sec:mlkws}
\subsection{Direction-Aware Enhancement Overview}\label{ov}
In this section, we review the task of separating the target speaker from a multi-channel speech mixture by making use of target speaker's direction information. Previous work in \cite{chen2018multi,gu2019neural,bahmaninezhad2019comprehensive,Gu20} have proposed to leverage a proper designed directional feature of the target speaker to perform the target speaker separation. The work in both \cite{bahmaninezhad2019comprehensive} and \cite{Gu20} implemented the enhancement network by a dilated convolutional neural network (CNN) similar as conv-TasNet \cite{luo2019convtasnet} but through a short-time Fourier transform (STFT) for signal encoding. Such network structure supports a long reception field to capture more sufficient contextual information and is thus adopted in our model.

Similar as the diagram in Figure \ref{fig:mlkws}, the direction-aware enhancement (DAE) framework starts from an encoder that maps the multi-channel input waveforms to complex spectrograms by a STFT 1-D convolution layer. Based on the complex spectrograms, the single-channel spectral feature, logarithm power spectrum (LPS) and multi-channel spatial features are extracted. A reference channel, e.g. the first channel complex spectrogram $\mathbf{Y}_1$, is used to compute LPS by $\text{LPS} = \log(|\mathbf{Y}_1|^2) \in \mathbb{R}^{T\times F}$, where $T$ and $F$ are the total number of frames and frequency bands of the complex spectrogram, respectively. One of the spatial features, IPD, is computed by the phase difference between channels of complex spectrograms as:
\begin{equation}
\text{IPD}^{(m)}(t,f)=\angle\mathbf{Y}_{m_1}(t,f)-\angle\mathbf{Y}_{m_2}(t,f)
\label{eq:ipd}
\end{equation}
where $m_1$ and $m_2$ are two microphones of the $m$-th microphone pair out of $M$ selected microphone pairs.
A directional feature (DF) is incorporated as a target speaker bias. This feature was originally introduced in \cite{chen2018multi}, which computes the averaged cosine distance between the target speaker steering vector and IPD on all selected microphone pairs as
\vspace{-0.15cm}
\begin{equation}
\vspace{-0.2cm}
\text{d}_\theta(t, f)=\overset{M}{\underset{m=1}{\sum}}
\left <\mathbf{e}^{\angle\mathbf{v}_{\theta}^{(m)}(f)},
\mathbf{e}^{\text{IPD}^{(m)}(t,f)} \right >
\label{eq4}
\end{equation}
where $\angle\mathbf{v}_{\theta}^{(m)}(f):=2\pi f\varDelta^{(m)}  \cos{\theta^{(m)}} / c$ is phase of the steering vector for target speaker from $\theta$ at frequency \emph{f} with respect to $m$-th microphone pair, $\varDelta^{(m)}$ is the distance between the $m$-th microphone pair, $c$ is the sound velocity, and vector $\mathbf{e}^{(\cdot)}:= [\cos(\cdot), \sin(\cdot)]^T$. If the T-F bin $(t,f)$ is dominated by the source from $\theta$, then $d_{\theta}(t,f)$ will be close to 1, otherwise close to 0. As a result, $\text{d}_\theta(t,f)$ indicates if a speaker from a desired direction $\theta$ dominates in each T-F bin, which drives the network to extract the target speaker from the mixture.
All of the features above are then concatenated and passed to the enhancement blocks, which consist of stacked dilated convolutional layers with exponentially growing dilation factors \cite{luo2019convtasnet}. The predicted target speaker mask is multiplied by the complex spectrogram of reference channel $\mathbf{Y}_1$. At the end, an inverse STFT (iSTFT) 1-D convolution layer converts the estimated target speaker complex spectrogram to the waveform.

Furthermore, the scale-invariant signal-to-noise (SI-SNR) is used as the objective function to optimize the enhancement network which is defined as:
\vspace{-0.1cm}
\begin{equation}
    \text{SI-SNR}(\hat{\mathbf{x}}, \mathbf{x}):=10\log_{10}\frac
{\left\|\mathbf{x}_{\sf target}\right\|_{2}^{2}}
{\left\|\mathbf{e}_{\sf noise}\right\|_{2}^{2}}
\label{eq1}
\end{equation}
where $\mathbf{x}_{\sf target}=\left(\left<\hat{\mathbf{x}}, \mathbf{x}\right>\mathbf{x}\right)/\left\|\mathbf{x}\right\|_{2}^{2}$, $\mathbf{e}_{\sf noise}=\hat{\mathbf{x}}-\mathbf{x}_{\sf target}$, and $\hat{\mathbf{x}}$ and $\mathbf{x}$ are the estimated and reverberant target speech waveforms, respectively. The zero-mean normalization is applied to $\hat{\mathbf{x}}$ and $\mathbf{x}$ for scale invariance. This loss function has been proven superior to MSE loss in \cite{bahmaninezhad2019comprehensive}.

\subsection{Multi-Look Enhancement Network}\label{subsec:mlen}

The DAE model in Section \ref{ov} relies on the correct estimation of the desired speaker's DOA information. However, the target direction estimation is infeasible under noisy conditions, particularly when the interfering sources are competing talkers. The idea of ``multi-look direction'' has been applied to speech separation \cite{Ji20,Chen17b,Chen18a} and multi-channel acoustic model \cite{Sainath15,Sainath16, Sainath17}, respectively, where a small number of spatial look directions cover
all possible target speaker directions. Since beamforming shows its advantage for speech preservation through its linear spatial filter design and processing \cite{higuchi2016robust, heymann2016neural,erdogan2016improved, xiao2016study}, a set of beamformers of different main lobe directions is thus used for multi-look enhancement in \cite{Ji20,Chen17b,Chen18a}. Neural network based multi-look filtering in \cite{Sainath15,Sainath16,Sainath17} implicitly learns filters for enhancing sources from different spatial look directions and passes all the filtered signals to an acoustic model for joint training. The multi-look enhancement layers are not trained by enhancement loss in a supervised mode. Such multi-look learning is not well controllable and thereby is hard to enhance and reconstruct the target speaker waveform at any look direction. Based on the target speaker enhancement architecture in Section \ref{ov}, we present a novel supervised multi-look neural enhancement model.

As shown in Figure \ref{fig:mlkws}, a set of $K$ directions in the horizontal plane is sampled.
The azimuths of look directions $\Theta_{1,2,...,K}$ result in $K$ directional feature vectors $d(\Theta_k), k=1,2,...,K$.
Per discussion in Section \ref{ov}, the value of directional feature in a T-F bin is close to 1 if the source from the desired direction is dominant in this bin. Furthermore, this value decreases as the source deviates from the desired look direction. To be more specific, under the free field assumption, i.e. only direct path of the acoustic sound is considered, we have $\text{IPD}^{(m)}(t,f) \approx \angle\mathbf{v}_{\theta}^{(m)}(f)$,
assuming the T-F bin is occupied by a source from direction $\theta$.
Therefore, at such T-F bin the directional feature of look direction $\Theta_k, k = 1,2,...,K$ can be approximated by
\vspace{-0.3cm}
\begin{equation}
\vspace{-0.2cm}
\text{d}_{\Theta_k}(t, f)\approx \overset{M}{\underset{m=1}{\sum}}
\left <\mathbf{e}^{\angle\mathbf{v}_{\Theta_k}^{(m)}(f)},
\mathbf{e}^{\angle\mathbf{v}_{\theta}^{(m)}(f)} \right >
\label{eq_dml}
\end{equation}
Obviously, $\text{d}_{\Theta_k}(t, f)$ is determined by the actual source direction $\theta$ and look direction $\Theta_k$. As a result, $\text{d}_{\Theta_k}$ for those T-F bins dominated by the source that is closest to the look direction $\Theta_k$ will be larger than that for other T-F bins. Such directional features enable the network to predict $K$ output channels $\hat{\mathbf{x}}^k, k=1,2,...,K$, corresponding to the closest source to each look direction, respectively. The supervised assignment are expressed as $\mathbf{x}^{k}=\mathbf{x}_{\tilde{k}}$
with
\vspace{-0.1cm}
\begin{equation}
\vspace{-0.1cm}
    \tilde{k}=\mathop{\arg\min}_{j}|\Theta_k - \theta_j|,
\end{equation}
where $\theta_j$ is the DOA of source $x_j$ in the mixture waveform, and $j=1,2,...,N$

In other words, the multi-look enhancement network simultaneously predicts the most nearby source for each look direction. The loss function thereby becomes
\vspace{-0.1cm}
\begin{equation}
\vspace{-0.1cm}
\mathcal{L}=
\overset{K}{\underset{k=1}{\sum}}
\text{SI-SNR}(\hat{\mathbf{x}}^k, \mathbf{x}^k)
\label{eq8}
\end{equation}

\begin{figure}[t]
  \centering
  \includegraphics[width=80mm, height=65mm]{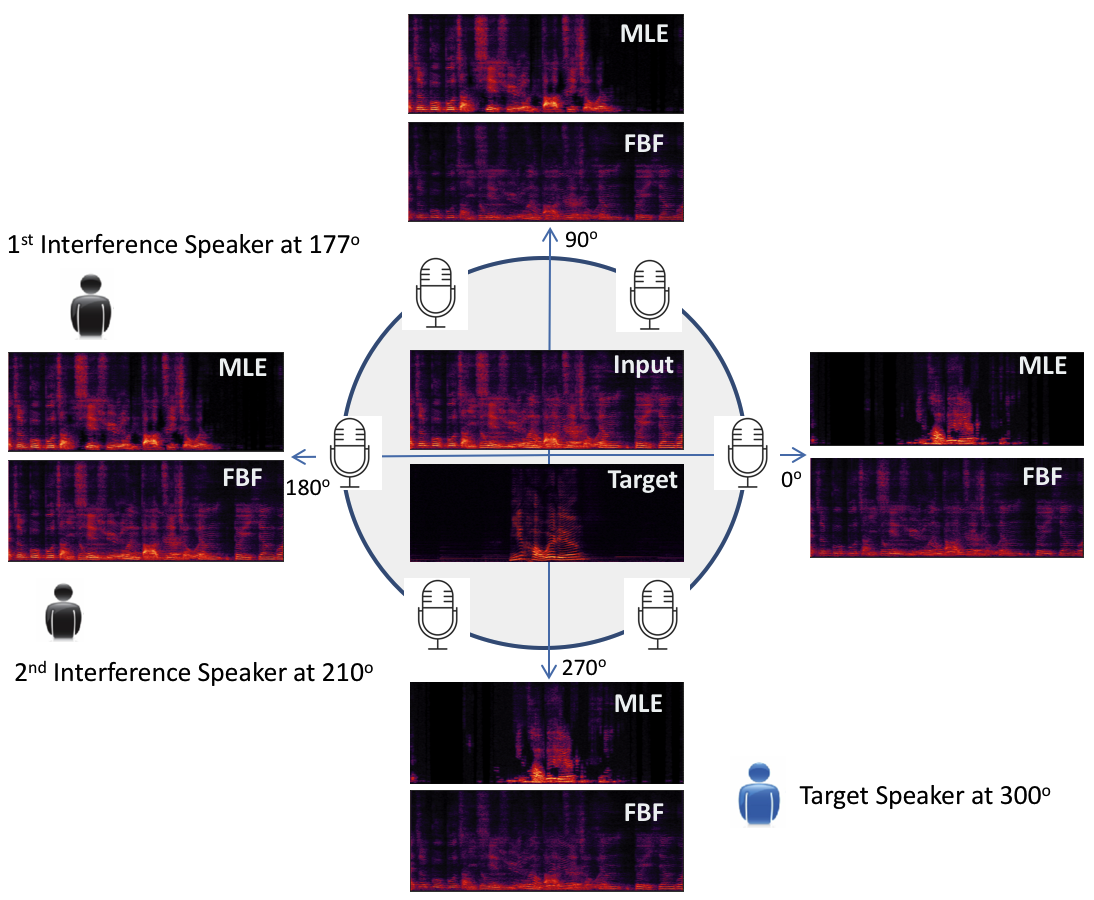}
  \caption{Example of MLENet outputs on 4 look directions and comparison to FBF. The circular microphone array records sounds from one target speaker and two speakers of interference. The two spectrograms in the middle are input mixture and target speaker reference, respectively. }
\vskip -0.4cm
  \label{demo}
\end{figure}

In our experiments, based on a uniform circular array of 6 microphones, we empirically use 4 look directions, targeting at $0^{\circ}$, $90^{\circ}$, $180^{\circ}$ and $270^{\circ}$, respectively, to cover the whole horizontal plane of $360^{\circ}$. An output example of multi-look enhancement network (MLENet) and fixed beamformers (FBF) is shown in Figure \ref{demo}. MLENet enhances the target speaker at the look direction $0^{\circ}$ and $270^{\circ}$, respectively, as the target speaker is closer to the two directions than other two speakers. Due to the capability of fixed beamformer, interference speakers are not well attenuated in any look direction. Based on this example, we emphasize that the target speaker may not be predicted which happens on the 2nd speaker of interference in this case. We call it ``off-target'' as the 2nd speaker of interference is not closer to any look direction compared to other speakers. We will discuss improved solutions in Section \ref{jt} and \ref{rst}.

\subsection{Joint Training with KWS Model}\label{jt}
The more look directions we have, the more likely the target speaker is closer to at least one look direction compared to other speakers of interference and thus exists in the output channels. Unfortunately, this may complicate the KWS model by performing keyword detection on each of the output channels. Therefore, we propose to integrate the output channels from multiple look directions into a single KWS model by jointly training the KWS and MLENet. Since the space resolution of the sampled look directions is not necessarily sufficient enough to cover the target direction, the mismatch between the target speaker direction and the look-direction causes either speech distortion in the output or even ``off-target'' in the output channels. An extra channel from one reference microphone is thereby leveraged to preserve target speech quality in those extremely difficult scenarios. A schematic diagram of the proposed system is shown in Figure \ref{fig:mlkws}.

Inspired by the application of attention mechanism in speech recognition \cite{SKim17}, speaker verification \cite{Chowdhury17} and single channel keyword spotting \cite{Shan18}, following \cite{Ji20} we incorporate a soft self-attention for projecting $K+1$ channels' fbank feature vectors to one channel, so that KWS still takes one channel input vector similarly as the baseline single channel model. For each time-step, we compute a $K+1$ dimensional attention weight vector $\bm{\alpha}$ for input fbank feature vectors $\mathbf{z} = \left[\mathbf{z}_1, \mathbf{z}_2, \dots, \mathbf{z}_K, \mathbf{z}_{K+1}\right]$ as:
\begin{align}
    e_i &= v^T tanh(W\mathbf{z}_i + b) \\
    \alpha_i & = \frac{exp(e_i)}{\sum_{k=1}^{K+1}exp(e_k)}
\end{align}
where a shared-parameter non-linear attention with the same $W$, $b$ and $v$ is used for each channel $i$ of all $K + 1$ channels. $\mathbf{z}$ is a 5-channel input fbank feature tensor in our implementation, corresponding to 4 multi-look enhanced signals and 1 reference microphone signal. $W$ is a $128\times D$ weight matrix where $D$ is the input feature size defined in Section \ref{subsec:kws}, $b$ is a 128-dimension bias vector, and $v$ is a 128-dimension vector.
A weighted sum of the multi-channel inputs is computed as
\begin{equation}
    \hat{\mathbf{z}} = \sum_{i=1}^{K+1} \alpha_i \mathbf{z}_i
\end{equation}
The KWS network and MLENet are then jointly optimized towards the improved keyword recognition accuracy.

\section{Experiments}\label{exp}
\subsection{KWS Pre-training}\label{subsec:kws}
Our baseline KWS model uses a Limited weight sharing (LWS) scheme based CNN \cite{Hamid14}, which consists of a convolutional layer (eight $4\times1$ non-overlapping kernels for eight different regions of frequency bands), a pooling layer, three fully connected layers each with 384 units, a fully connected layer with 128 units, and a softmax layer. 40 dimensional log-mel filterbank features are computed every 25ms with a 10ms frame shift and their delta and delta-delta features are appended. At each frame, we stack 10 frames to the left and 5 frames to the right as the input feature to the convolutional layer. The baseline KWS model is trained on large internal training
sets to detect the keyword ``ni-hao-wei-ling'' in Mandarin. The interested reader is referred to \cite{MYu18,Chen14} for more details on modeling and decoding.

A 200-hour keyword specific data set was used as positive training examples. It is from 337 human speakers and includes 45K utterances from headset recordings (relatively clean data) and 179K utterances from a distant microphone (far-field noisy data).
A 139-hour dataset of 100K negative examples from a Mandarin speech database served as negative training examples.

\subsection{MLENet Pre-training}
The window size is 32 ms and the hop size is 16 ms. We apply 512-point
FFT to extract 257-dimensional LPS and spatial
features (IPD and DF) for MLENet training. IPDs are extracted from 6 microphone pairs, ($0^{\circ}$, $180^{\circ}$), ($60^{\circ}$, $240^{\circ}$), ($120^{\circ}$, $300^{\circ}$), ($0^{\circ}$, $60^{\circ}$), ($120^{\circ}$, $180^{\circ}$) and ($240^{\circ}$, $300^{\circ}$), where the angle values indicate the microphone positions illustrated in Figure \ref{demo}.
The design of enhancement blocks follows \cite{luo2019convtasnet}, including 4 times' repeats of 8 convolutional blocks with dilation factors $1, 2, 4,..., 2^7$.

We simulate a multi-channel dataset of reverberant mixtures with up to three speakers in each utterance by AISHELL-2 corpus \cite{Du2018}.
The room simulator based on the image method \cite{Allen79} generates 10K rooms with random room characteristics, speaker and array locations. A 6-element uniform circular array of radius 0.035~m is simulated as the receiver. The corresponding room sizes (length$\times$width$\times$height) range from $3m\times3m\times2.5m$ to $8m\times10m\times6m$. The reverberation time $T_{60}$s range from 0 to 600 ms, with an average $T_{60}$ of 300 ms. The simulated room impulse responses (RIRs) are randomly selected for creating waveform mixtures of random number of speakers.
The signal-to-interference-ratio (SIR) of one speaker with respect to other overlapped speakers is randomly drawn from -12 dB to 12dB. Extra environmental noise sources are mixed with the simulated utterances with signal-to-noise-ratio (SNR) randomly sampled from 12dB to 30dB. We generate 95K utterances and 3K utterances for training and validation, respectively. 

\begin{table}[t]
\centering
\caption{\label{se_eval} {\it SI-SNR (dB) evaluation in different SIR conditions}}
\vspace{-0.2cm}
\begin{tabular}{l|ccc}\toprule
 \midrule  Front-end  &  $<$ 6dB &  $>=$ 6dB & w/o Intf. \\
  \midrule raw Input & -9.46 & 7.78 & 17.66 \\
   DAE & 4.01 & 16.33 & 22.42 \\
   MLENet (pre-train)  &  0.87 & 14.59 & 21.72 \\
   MLENet (joint train)   & 4.50 & 14.97 & 24.53\\
\bottomrule
\end{tabular}
\vspace{-0.6cm}
\end{table}

\subsection{Multi-Look KWS Joint Training}
The pre-trained MLENet and KWS network are jointly finetuned by the mixture waveforms in 6-microphone circular array. Similar as the way we create the dataset for pre-training MLENet, a target speaker of keyword utterance is mixed with up to two speakers of interference in each mixture. The SIR at overlapped periods is randomly drawn from -12 dB to 30dB. The environmental noise is added with SNR randomly sampled from 12dB to 30dB. A total number of 154K and 15K simulated utterances served as positive samples for training and evaluation, respectively. 28K and 47K (about 60 hours) negative samples are used for training and false alarm test, respectively.

\subsection{Results and Discussion}\label{rst}
We first prove the effectiveness of the proposed MLENet on the task of target speaker enhancement. The 15K evaluation mixture utterances of target spoken wake-up words, speakers of interference and environmental noises  are used for SI-SNR evaluation in Table \ref{se_eval}. The evaluations are grouped to three conditions, multi-talker (up to two speakers of interference) with SIR below 6dB, multi-talker with SIR above 6dB and none interference conditions, respectively. Environmental noises are applied to all three far-field sets with SNR above 12dB. Direction-aware enhancement described in Section \ref{ov} performs very well with the oracle DOA of target speaker and thus serves as an up-bound for the multi-look enhancement. For the MLENet's 4 output channels, the best SI-SNR is presented in this table as the target speaker is not predicted in a certain output channel. The pre-trained MLENet performs reasonably well in all conditions. Due to the discussed speech distortion and ``off-target'' issues, there are about 3dB, 2dB and less than 1dB gaps to the DAE with oracle DOA in three categories, respectively. The last row of Table \ref{se_eval} shows that MLENet in the jointly trained multi-look KWS model significantly improves the its robustness and reduces the target speech distortion.

\begin{figure}[t]
  \hspace*{-0.4cm}
  \includegraphics[width=92mm, height=32mm]{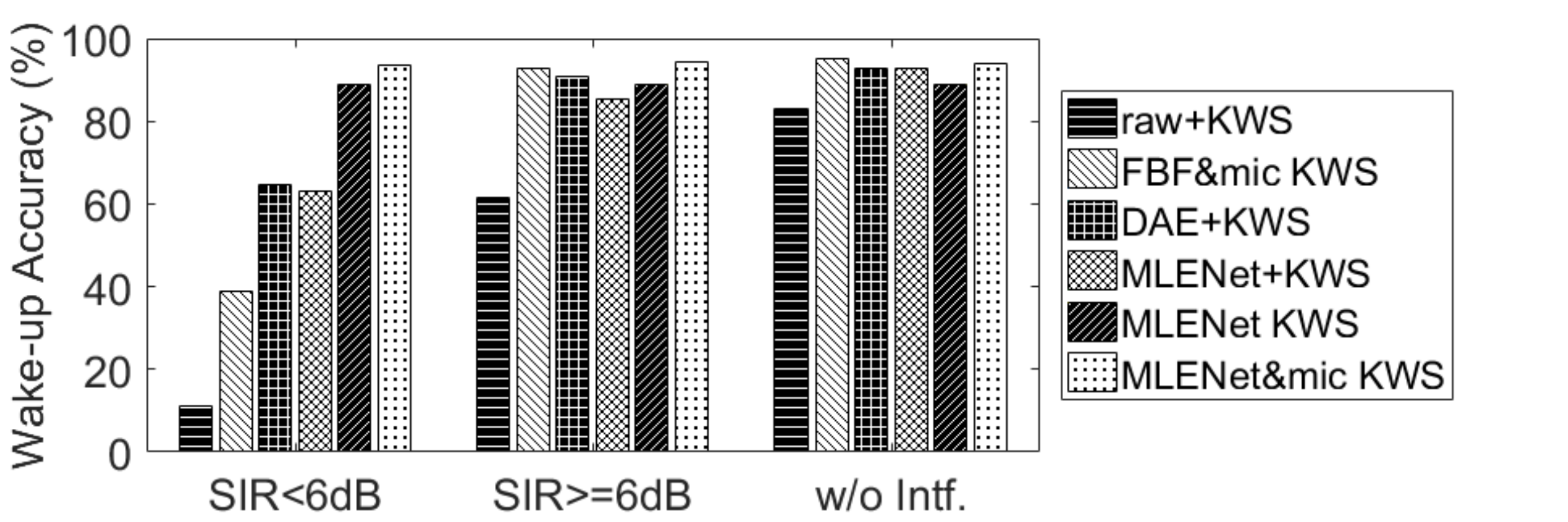}
  \caption{Wake-up accuracy with one time false alarm in 12 hours. The evaluated methods are baseline KWS with no front-end processing (raw+KWS), fixed beamformer based multi-beam KWS with an extra microphone channel in the KWS input\cite{Ji20} (FBF\&mic KWS), front-end DAE with oracle target DOA followed by baseline KWS, front-end MLENet followed by baseline KWS on each individual MLENet output channel (MLENet+KWS), multi-look KWS without using extra microphone channel (MLENet KWS), and multi-look KWS by incorporating an extra microphone channel (MLENet\&mic KWS). }
\vspace{-0.5cm}
  \label{kwscomp}
\end{figure}

Figure \ref{kwscomp} shows KWS performance measured by wake-up accuracy under the setup that up to one time false alarm triggered in 12 hours' exposure to continuous speech, TV, and a variety of noises. Compared to finding the equal error rate from the receiver operating characteristic curve, this evaluation metric conforms better to industry assessment.
Compared to the baseline KWS (raw+KWS), the improvement by MLENet front-end processing (MLENet+KWS) are quite significant, especially in multi-talker conditions. We can see that MLENet achieves comparable wake-up accuracy as DAE with oracle DOA. The jointly trained multi-look KWS (MLENet KWS) shows enhanced performance compared to KWS with front-end MLENet processing (MLENet+KWS). Furthermore, MLENet\&mic KWS outperforms the one without using microphone channel (MLENet KWS), indicating the contribution from the microphone channel for handling target speaker distortion and ``off-target'' cases. Although beamformer based multi-beam KWS achieves fairly good performance, particularly in moderate to high SIR and SNR conditions, the proposed multi-look KWS proves a great advantage in low SIR conditions. The wake-up accuracies for MLENet\&mic KWS are 93.4\%, 94.5\% and 94.0\%, showing promising steady performance in all three conditions, respectively. We counted the percentage of ``off-target'' cases in the two evaluation categories with speakers of interference (SIR$<$6dB \& SIR$ >=$ 6dB) where the target speaker may be absent in the output channels. The value is about 9\% in each category. By looking at the multi-look KWS accuracy that is around 94\%, it proves that the extra microphone channel and end-to-end joint training improve the robustness of MLENet and the whole system.

\section{Conclusions}\label{con}
In this paper, we proposed a multi-look enhancement network (MLENet), which simultaneously enhances the acoustic sources from multiple look directions. The key idea is to utilize a directional feature on multiple look directions as the input features. Such directional features solves the output assignment difficulty and enables the supervised training of MLENet.
The formulation of multi-look enhancement in a neural network allows us to perform end-to-end training. Experimental results show that the proposed approach significantly outperforms the baseline KWS system and the beamforming based multi-beam KWS system. We observe that MLENet will be easily generalized to work with speaker verification and speech recognition in future work.

\newpage
\bibliographystyle{IEEEtran}
\bibliography{mybib}

\begin{thebibliography}{10}
\providecommand{\url}[1]{#1}
\csname url@samestyle\endcsname
\providecommand{\newblock}{\relax}
\providecommand{\bibinfo}[2]{#2}
\providecommand{\BIBentrySTDinterwordspacing}{\spaceskip=0pt\relax}
\providecommand{\BIBentryALTinterwordstretchfactor}{4}
\providecommand{\BIBentryALTinterwordspacing}{\spaceskip=\fontdimen2\font plus
\BIBentryALTinterwordstretchfactor\fontdimen3\font minus
  \fontdimen4\font\relax}
\providecommand{\BIBforeignlanguage}[2]{{%
\expandafter\ifx\csname l@#1\endcsname\relax
\typeout{** WARNING: IEEEtran.bst: No hyphenation pattern has been}%
\typeout{** loaded for the language `#1'. Using the pattern for}%
\typeout{** the default language instead.}%
\else
\language=\csname l@#1\endcsname
\fi
#2}}
\providecommand{\BIBdecl}{\relax}
\BIBdecl

\bibitem{Rohlicek89}
J.~Rohlicek, W.~Russell, S.~Roukos, and H.~Gish, ``Continuous hidden markov
  modeling for speaker-independent word spotting,'' in \emph{the Proceedings of
  ICASSP}, 1989, pp. 627--630.

\bibitem{Ephraim85}
Y.~Ephraim and D.~Malah, ``Speech enhancement using a minimum mean-square error
  log-spectral amplitude estimator,'' \emph{IEEE Transactions on Acoustics,
  Speech, and Signal Processing}, vol.~33, no.~2, pp. 443--445, 1985.

\bibitem{Wang06}
D.~Wang and G.~Brown, \emph{Computational Auditory Scene Analysis: Principles,
  Algorithms, and Applications}.\hskip 1em plus 0.5em minus 0.4em\relax
  Wiley-IEEE Press, 2006.

\bibitem{Wang14}
Y.~Wang, A.~Narayanan, and D.~Wang, ``On training targets for supervised speech
  separation,'' \emph{IEEE Transactions on Audio, Speech, and Language
  Processing}, vol.~22, no.~12, pp. 1849--1858, 2014.

\bibitem{Xu15}
Y.~Xu, J.~Du, L.~Dai, and C.~Lee, ``A regression approach to speech enhancement
  based on deep neural networks,'' \emph{IEEE/ACM Transactions on Audio,
  Speech, and Language Processing}, vol.~23, no.~1, pp. 1849--1858, 2015.

\bibitem{Weninger15}
F.~Weninger, H.~Erdogan, S.~Watanabe, E.~Vincent, J.~L. Roux, J.~R. Hershey,
  and B.~Schuller, ``Speech enhancement with lstm recurrent neural networks and
  its application to noise-robust asr,'' \emph{in Latent Variable Analysis and
  Signal Separation, Springer}, pp. 91--99, 2015.

\bibitem{Hershey16}
J.~R. Hershey, Z.~Chen, J.~L. Roux, and S.~Watanabe, ``Deep clustering:
  Discriminative embeddings for segmentation and separation,'' in \emph{the
  Proceedings of International Conference on Acoustics, Speech and Signal
  Processing (ICASSP). IEEE}, 2016, pp. 31--35.

\bibitem{Chen17c}
Z.~Chen, Y.~Luo, and N.~Mesgarani, ``Deep attractor network for
  single-microphone speaker separation.'' pp. 246--250.

\bibitem{Kolbk17}
M.~Kolbak, D.~Yu, Z.-H. Tan, and J.~Jensen, ``Multitalker speech separation
  with utterance-level permutation invariant training of deep recurrent neural
  networks,'' \emph{IEEE/ACM Transactions on Audio, Speech and Language
  Processing}, vol.~25, no.~10, pp. 1901--1913, 2017.

\bibitem{MYu18}
M.~Yu, X.~Ji, Y.~Gao, L.~Chen, J.~Chen, J.~Zheng, D.~Su, and D.~Yu,
  ``Text-dependent speech enhancement for small-footprint robust keyword
  detection,'' in \emph{Interspeech}, 2018, pp. 2613--2617.

\bibitem{chen2018multi}
Z.~Chen, X.~Xiao, T.~Yoshioka, H.~Erdogan, J.~Li, and Y.~Gong, ``Multi-channel
  overlapped speech recognition with location guided speech extraction
  network,'' in \emph{IEEE Spoken Language Technology Workshop (SLT)}, 2018,
  pp. 558--565.

\bibitem{wang2018multi}
Z.-Q. Wang, J.~Le~Roux, and J.~R. Hershey, ``Multi-channel deep clustering:
  Discriminative spectral and spatial embeddings for speaker-independent speech
  separation,'' in \emph{IEEE International Conference on Acoustics, Speech and
  Signal Processing (ICASSP)}, 2018, pp. 1--5.

\bibitem{lianwu2019multi}
L.~Chen, M.~Yu, D.~Su, and D.~Yu, ``Multi-band pit and model integration for
  improved multi-channel speech separation,'' in \emph{IEEE International
  Conference on Acoustics, Speech and Signal Processing (ICASSP)}.\hskip 1em
  plus 0.5em minus 0.4em\relax IEEE, 2019.

\bibitem{wang2019combining}
Z.~Wang and D.~Wang, ``Combining spectral and spatial features for deep
  learning based blind speaker separation,'' \emph{IEEE/ACM Transactions on
  Audio, Speech, and Language Processing}, vol.~27, no.~2, pp. 457--468, 2019.

\bibitem{gu2019neural}
R.~Gu, L.~Chen, S.-X. Zhang, J.~Zheng, Y.~Xu, M.~Yu, D.~Su, Y.~Zou, and D.~Yu,
  ``Neural spatial filter: Target speaker speech separation assisted with
  directional information,'' in \emph{Proc. Interspeech}, 2019.

\bibitem{bahmaninezhad2019comprehensive}
F.~Bahmaninezhad, J.~Wu, R.~Gu, S.-X. Zhang, Y.~Xu, M.~Yu, and D.~Yu, ``A
  comprehensive study of speech separation: spectrogram vs waveform
  separation,'' \emph{Proc. Interspeech}, 2019.

\bibitem{Ji20}
X.~Ji, M.~Yu, J.~Chen, J.~Zheng, D.~Su, and D.~Yu, ``Integration of multi-look
  beamformers for multi-channel keyword spotting,'' in \emph{the Proceedings of
  ICASSP}, 2020, pp. 7464--7468.

\bibitem{Gu20}
R.~Gu, S.-X. Zhang, Y.~Xu, L.~Chen, Y.~Zou, , and D.~Yu, ``Multi-modal
  multi-channel target speech separation,'' \emph{IEEE Journal of Selcted
  Topics in Signal Processing,}, 2020.

\bibitem{luo2019convtasnet}
Y.~{Luo} and N.~{Mesgarani}, ``Conv-tasnet: Surpassing ideal time–frequency
  magnitude masking for speech separation,'' \emph{IEEE/ACM Transactions on
  Audio, Speech, and Language Processing}, vol.~27, no.~8, pp. 1256--1266, Aug
  2019.

\bibitem{Chen17b}
Z.~Chen, J.~Li, X.~Xiao, T.~Yoshioka, H.~Wang, Z.~Wang, and Y.~Gong, ``Cracking
  the cocktail party problem by multi-beam deep attractor network,'' in
  \emph{IEEE Workshop on ASRU}, 2017.

\bibitem{Chen18a}
Z.~Chen, T.~Yoshioka, X.~Xiao, J.~Li, M.~L. Seltzer, and Y.~Gong, ``Efficient
  integration of fixed beamformers and speech separation networks for
  multi-channel far-field speech separation,'' in \emph{the Proceedings of
  ICASSP}, 2018.

\bibitem{Sainath15}
T.~N. Sainath, R.~J. Weiss, K.~W. Wilson, A.~Senior, and O.~Vinyals, ``Learning
  the speech front-end with raw waveform cldnns,'' in \emph{Proc. Interspeech},
  2015.

\bibitem{Sainath16}
T.~N. Sainath, R.~J. Weiss, K.~W. Wilson, A.~Narayanan, and M.~Bacchiani,
  ``Factored spatial and spectral multichannel raw waveform cldnns,'' in
  \emph{IEEE International Conference on Acoustics, Speech and Signal
  Processing (ICASSP)}.\hskip 1em plus 0.5em minus 0.4em\relax IEEE, 2016.

\bibitem{Sainath17}
T.~N. Sainath, R.~J. Weiss, K.~W. Wilson, B.~Li, A.~Narayanan, E.~Variani,
  M.~Bacchiani, I.~Shafran, A.~Senior, K.~Chin, A.~Misra, and C.~Kim,
  ``Multichannel signal processing with deep neural networks for automatic
  speech recognition,'' \emph{IEEE/ACM Transactions on Audio, Speech, and
  Language Processing}, 2017.

\bibitem{higuchi2016robust}
T.~Higuchi, N.~Ito, T.~Yoshioka, and T.~Nakatani, ``Robust mvdr beamforming
  using time-frequency masks for online/offline asr in noise,'' in
  \emph{ICASSP}.\hskip 1em plus 0.5em minus 0.4em\relax IEEE, 2016, pp.
  5210--5214.

\bibitem{heymann2016neural}
J.~Heymann, L.~Drude, and R.~Haeb-Umbach, ``Neural network based spectral mask
  estimation for acoustic beamforming,'' in \emph{ICASSP}.\hskip 1em plus 0.5em
  minus 0.4em\relax IEEE, 2016, pp. 196--200.

\bibitem{erdogan2016improved}
H.~Erdogan, J.~R. Hershey, S.~Watanabe, M.~I. Mandel, and J.~Le~Roux,
  ``Improved mvdr beamforming using single-channel mask prediction networks.''
  in \emph{Interspeech}, 2016, pp. 1981--1985.

\bibitem{xiao2016study}
X.~Xiao, C.~Xu, Z.~Zhang, S.~Zhao, S.~Sun, S.~Watanabe, L.~Wang, L.~Xie, D.~L.
  Jones, E.~S. Chng \emph{et~al.}, ``A study of learning based beamforming
  methods for speech recognition,'' in \emph{CHiME 2016 workshop}, 2016, pp.
  26--31.

\bibitem{SKim17}
S.~Kim and I.~Lane, ``End-to-end speech recognition with auditory attention for
  multi-microphone distance speech recognition,'' in \emph{Interspeech}, 2017,
  pp. 3867--3871.

\bibitem{Chowdhury17}
F.~Chowdhury, Q.~Wang, I.~L. Moreno, and L.~Wan, ``Attention based models for
  text-dependent speaker verification,'' \emph{arXiv preprint
  arXiv:1710.10470}, 2017.

\bibitem{Shan18}
C.~Shan, J.~Zhang, Y.~Wang, and L.~Xie, ``Attention-based end-to-end models for
  small-footprint keyword spotting,'' in \emph{Interspeech}, 2018, pp.
  1571--1575.

\bibitem{Hamid14}
O.~Abdel-Hamid, A.-R. Mohamed, H.~Jiang, L.~Deng, G.~Penn, and D.~Yu,
  ``Convolutional neural networks for speech recognition,'' \emph{IEEE/ACM
  Trans. Audio, Speech, Lang. Process.}, vol.~22, no.~10, pp. 1533--1545, 2014.

\bibitem{Chen14}
G.~Chen, C.~Parada, and G.~Heigold, ``Small-footprint keyword spotting using
  deep neural networks,'' in \emph{the Proceedings of ICASSP}, 2014, pp.
  4087--4091.

\bibitem{Du2018}
J.~Du, X.~Na, X.~Liu, and H.~Bu, ``Aishell-2: Transforming mandarin asr
  research into industrial scale,'' \emph{arXiv:1808.10583}, 2018.

\bibitem{Allen79}
J.~B. Allen and D.~A. Berkley, ``Image method for efficiently simulation
  room-small acoustic,'' \emph{Journal of the Acoustical Society of America},
  vol.~65, no.~4, pp. 943--950, 1979.

\end{thebibliography}

\end{document}